\documentclass[%
 reprint,
 twocolumn,
 nofootinbib,
 superscriptaddress,
 showpacs,
 preprintnumbers,
 amsmath,
 amssymb,
 aps,
 prd,
 floatfix,
]{revtex4-1}

\usepackage{latexsym} 	
\usepackage{amssymb} 
\usepackage{amsbsy}
\usepackage{amsmath}
\usepackage{epsfig}     
\usepackage{graphics,color}
\usepackage{multirow}
\usepackage{hyperref}       

\newcommand{\be}{\begin{equation}}
\newcommand{\ee}{\end{equation}}
\newcommand{\bea}{\begin{eqnarray}}
\newcommand{\eea}{\end{eqnarray}}
\newcommand{\bean}{\begin{eqnarray*}}
\newcommand{\eean}{\end{eqnarray*}}
\newcommand{\bit}{\begin{itemize}}
\newcommand{\eit}{\end{itemize}}
\newcommand{\eps}{\epsilon}

\newcommand{\nn}{\nonumber}
\newcommand{\half}{\frac{1}{2}}
\newcommand{\id}{1\!\!1}

\newcommand{\qqquad}{\qquad\qquad}

\newcommand{\ket}{\rangle}
\newcommand{\bra}{\langle}

\newcommand{\xv}{\mathbf{x}}
\newcommand{\yv}{\mathbf{y}}
\newcommand{\cB}{{\cal B}}
\newcommand{\cL}{{\cal L}}
\newcommand{\cN}{{\cal N}}
\newcommand{\cO}{{\cal O}}
\newcommand{\Var}{{\rm Var}}

\begin{document}

\title{Stochastic weight matrix dynamics during learning and Dyson Brownian motion}

\author{Gert Aarts}
\email{g.aarts@swansea.ac.uk}
\affiliation{Department of Physics, Swansea University, Swansea SA2 8PP, United Kingdom}

\author{Biagio Lucini}
\email{b.lucini@swansea.ac.uk}
\affiliation{Department of Mathematics, Swansea University (Bay Campus), Swansea, SA1 8EN, United Kingdom}

\author{Chanju Park}
\email{chanju.b.park@gmail.com}
\affiliation{Department of Physics, Swansea University, Swansea SA2 8PP, United Kingdom}

\date{July 23, 2024. Revised: October 14, 2024}

\begin{abstract}
We demonstrate that the update of weight matrices in learning algorithms can be described in the framework of Dyson Brownian motion, thereby inheriting many features of random matrix theory. We relate the level of stochasticity to the ratio of the learning rate and the mini-batch size, providing more robust evidence to a previously conjectured scaling relationship. We discuss universal and non-universal features in the resulting Coulomb gas distribution and identify the Wigner surmise and Wigner semicircle explicitly in a teacher-student model and in the (near-)solvable case of the Gaussian restricted Boltzmann machine. 
\end{abstract}


\maketitle


\section{Introduction}
 \label{sec:intro}

Recent years have seen a dramatic increase of the use of machine learning (ML) in the fundamental sciences, with the adoption and development of many ML applications to improve and speed-up scientific analysis \cite{Carleo_2019}. Reversely there is a growing trend to use the methodology of (theoretical) physics to understand ML algorithms, viewing these as acting on systems with many fluctuating degrees of freedom and hence employing the analogy with statistical physics. In this paper we explore this second direction and argue that learning can be formulated in the framework of Dyson Brownian motion \cite{Dyson-4}, thereby inheriting many features of random matrix theory (RMT) \cite{Wigner-1,Wigner-2,Dyson-1,Dyson-2,Dyson-3,Meh2004}. 

In general terms, ML algorithms aim to minimise some cost function by applying stochastic gradient descent (SGD), or variations thereof, to weight matrices $W$ defined inside the architecture. Stochastic updates of matrices immediately establishes the link with Dyson Brownian motion, which exactly describes those, and yields an equation for the dynamics of the eigenvalues of the symmetric combination $X=W^TW$. This dynamics contains universal aspects, e.g.\ eigenvalue repulsion due to an induced Coulomb term, as reflected in the Wigner surmise, as well as non-universal aspects, related to details of the gradient of the loss function. In many algorithms, stochasticity is inherently present due sampling and mini-batch updates. We demonstrate that the distribution of eigenvalues depends on the ratio of the learning rate $\alpha$ and mini-batch size $|\cB|$, and not on these quantities separately. This {\em linear scaling rule} has been observed previously \cite{Goyal-1,Smith-1,Smith-2, Smith-3} 
(see also Ref.\ \cite{DBLP:journals/corr/abs-1710-11029}),
 but we demonstrate that it is a direct consequence of stochastic matrix dynamics and Dyson Brownian motion. An interesting corollary is that there is no simple limit (e.g.\  $\alpha\to 0$, $1/|\cB|\to 0$) in which SGD reduces to a stochastic differential equation (SDE) in continuous time \cite{mandt2015,li2017}. Again, this has been noted before \cite{yaida2018}, but we emphasise the viewpoint that a tunable ratio $\alpha/|\cB|$ in fact prevents an obvious SDE limit (in the limit  $\alpha/|\cB|\to 0$, the SGD update reduces to an ordinary differential equation for the eigenvalues).
The amount of stochasticity is directly proportional to $\alpha/|\cB|$ and therefore determines the strength of the eigenvalue repulsion. It hence sets a fundamental limit on the accuracy of learning, but reversely a larger level of stochasticity leads to better generalisation, as it avoids overfitting \cite{Goyal-1,Smith-1,Smith-2}. These concepts find a natural place in the framework of Dyson Brownian motion, as we will demonstrate.

This paper is organised as follows. After setting up the general framework in Sec.\ \ref{sec:weight}, we demonstrate our arguments explicitly in two cases, namely the Gaussian restricted Boltzmann machine (RBM) and a simple teacher-student model in Sec.\ \ref{sec:apply}. The conclusions are summarised in Sec.\ \ref{sec:con}, along with directions for future research. App.~\ref{app:dyson} contains a brief summary of Dyson Brownian motion, while Apps.\  \ref{app:MP}, \ref{app:scaling} and \ref{app:derivation} contain some further comments used in the main text. 

With relation to RMT, we note that Refs.\ \cite{Martin-2019,Baskerville} apply RMT to the distribution of eigenvalues of weight matrices in an empirical manner, but not in the framework of Dyson Brownian motion and the linear scaling rule.
In Ref.~\cite{Levi:2023qwg} it was shown that ``real-world data'' have many features consistent with RMT. Here we apply RMT to weight matrices, not to data.

\section{Stochastic weight matrix dynamics}
 \label{sec:weight}

\subsection{Stochastic updates}  

Let us consider some weight matrix $W$, which connects layers of nodes in a neural network or a restricted Boltzmann machine. During the learning stage, it is updated using stochastic gradient descent, or variations thereof, by subtracting the change in the loss function, $\cL[W]$,
\be
\label{eq:W}
W_{ij} \to W_{ij}' = W_{ij}+\delta W_{ij} = W_{ij}-\alpha \frac{\delta \cL}{\delta W_{ij}},
\ee
where $\alpha$ is the learning rate. 
This update is carried out using stochastically chosen mini-batches $\cB$ of size $|\cB|$, with a mini-batch average
\be
 \delta W_{ij}^\cB = \frac{1}{|\cB|} \sum_{b\in\cB}\delta W_{ij}^b.
 \ee
The entire data set from which the mini-batches are drawn is assumed to be sufficiently large to not lead to additional constraints. 
Since the mini-batch size is finite, $\delta W^\cB$ is a stochastic variable, with the size of the fluctuations set by the central limit theorem. We hence write the update as
\be
\label{eq:W2}
 \delta W_{ij}  = \delta W_{ij}^\cB + \frac{1}{\sqrt{|\cB|}}\sqrt{\Var(\delta W_{ij})}\, \eta_{ij},
  \ee
where the first term is the ``deterministic'' part of the update for a given mini-batch and the second part reflects stochastic fluctuations, whose magnitude decreases with increasing mini-batch size.
The noise $\eta_{ij}\sim\cN(0,1)$ is matrix-valued.
The variance is defined for each matrix element, i.e.,
 \be
 \Var(\delta W_{ij}) = \bra   \delta W_{ij}   \delta W_{ij}\ket - \bra\delta W_{ij}\ket\bra\delta W_{ij}\ket.
 \ee
 The update (\ref{eq:W2}) is applied to each matrix element and no summation over repeated indices is implied in this section. 
 In terms of the gradient of the loss function, this update reads
\be
\label{eq:Wex}
W_{ij}' = W_{ij} -\alpha\left(\frac{\delta\cal L}{\delta W_{ij} }\right)_{\cal B} +\frac{\alpha}{\sqrt{|{\cal B}|}}
\sqrt{\mbox{Var}\left(\frac{\delta\cal L}{\delta W_{ij} } \right)} \eta_{ij},
\ee
which makes the learning rate explicit.

Since $W$ is in general a rectangular $M\times N$ matrix, and the connection with random matrix theory is easier for symmetric matrices, we consider the symmetric combination
\be
\label{eq:X}
X =  W^TW.
\ee
Without loss of generality we take $N \leq M$; if this is not the case, simply exchange $W$ and $W ^T$. $X$ is a symmetric $N\times N$ matrix with $N$ semi-positive real eigenvalues $x_i =\xi_i^2$ ($i=1,...,N$), where $\xi_i$ are the singular values of $W$, obtained via the singular value decompositon $W = U\Xi V^T$.
The focus on the singular/eigenvalues also removes the redundancy of left/right rotations on the weight matrix. 
The update for $X$ follows from the one for $W$ as
\be
 X \to X' = X + \delta W^T W + W^T \delta W \equiv X+\delta X,
\ee
and for finite batch size, the update is stochastic, with
\be
\label{eq:deltaX}
X_{ij}\to X_{ij}' = X_{ij} +  \delta X_{ij}^{\cal B} + \frac{1}{\sqrt{|{\cal B|}}}\sqrt{\mbox{Var}(\delta X_{ij})} \eta_{ij},
\ee
as above, with symmetric noise in this case.

\subsection{Dyson Brownian motion}

The framework in which to consider stochastic matrix dynamics for a symmetric matrix $X$ is Dyson Brownian motion \cite{Dyson-4}, which is summarised in App.\ \ref{app:dyson}. The main feature is that the eigenvalues of $X$ not only evolve stochastically but also repel, due to an induced Coulomb term. The strength of the stochastic term and of the Coulomb term are related to the stochasticity in the original matrix equation for $X$. 

Starting from Eq.\ (\ref{eq:deltaX}), the eigenvalues $x_i$ of $X$ evolve according to -- see Eq.\ (\ref{eq:appx}) --
\be
\label{eq:x}
x_i\to x_i' = x_i + K_i +\sum_{j\neq i}\frac{g_i^2}{x_i-x_j} +\sqrt{2}g_i\eta_i,
\ee
where $K_i$ and $g_i$ are linked to the deterministic and stochastic terms in Eq.\ (\ref{eq:deltaX}), and again $\eta\sim\cN(0,1)$. The term with the summation is the Coulomb term, resulting in eigenvalue repulsion. 
The learning rate and batch size can be made explicit by writing, c.f.\ Eq.\ (\ref{eq:Wex}), 
\be
K_i = \alpha \tilde K_i,
\qqquad
g_i = \frac{\alpha}{\sqrt{|{\cal B|}}} \tilde g_i,
\ee
where $\tilde K_i$ and $\tilde g_i$ are related to the gradient of the loss function and its variance respectively (quantities with a tilde are independent of the learning rate and batch size at leading order).
The eigenvalue update then becomes
\be
\label{eq:xtilde}
x_i\to x_i' = x_i +  \alpha \tilde K_i + \frac{\alpha^2}{|{\cal B}|}\sum_{j\neq i}\frac{\tilde g_i^2}{x_i-x_j} + 
\frac{\alpha}{\sqrt{|{\cal B}|}}\sqrt{2}\tilde g_i\eta_i.
\ee
When trying to identify the learning rate or the inverse batch size with a stepsize $\eps$, we note here that the drift, including the Coulomb term, and the stochastic term do not scale in the standard (i.e.\ It\^o calculus) manner.
Indeed, the difficulties of going from stochastic gradient descent to a stochastic differential equation (SDE) \cite{mandt2015,li2017} are well known \cite{yaida2018}, with the update reducing to an ordinary differential equation in the limit that e.g.\ the learning rate goes to zero. 
However, as we will demonstrate shortly, the appearance of the learning rate and batch size as in Eq.\ (\ref{eq:xtilde}) naturally leads to the linear scaling rule, in which $\alpha/|\cB|$ is a parameter whose tunability can be exploited. An obvious SDE limit should therefore in fact not be expected. 

We will derive the linear scaling rule \cite{Smith-1,Smith-2} now, by considering the stationary distribution corresponding to the stochastic process (\ref{eq:x}). The associated Fokker-Planck equation for the distribution is given in App.\ \ref{app:dyson}. The stationary distribution is known as the Coulomb gas, which reads
\be
\label{eq:Cgas}
P_s(\{x_i\}) =  \frac{1}{Z}\prod_{i<j} \left|x_i-x_j\right| e^{-\sum_i V_i(x_i)/g_i^2},
\ee
with 
\be
Z = \int dx_1 \ldots dx_N\, P_s(\{x_i\}).
\ee
Here it is assumed that the drift can be derived from a separable potential,
\be
K_i(x_i) = -\frac{dV_i(x_i)}{dx_i}.
\ee
We can again make the learning rate and batch size explicit, by introducing
\be
V_i(x_i) = \alpha \tilde  V_i(x_i).
\ee
The combination in the exponent then reads
\be
\label{eq:Vi}
\frac{V_i(x_i)}{g_i^2} = \frac{1}{\alpha/|\cB|} \frac{\tilde V_i(x_i)}{\tilde g_i^2},
\ee
where the first factor on the RHS indicates universal scaling with $\alpha/|\cB|$ and the second factor depends on the details of the loss function.

If we assume that  the potentials $V_i$ have a minimum at $x_i = x_i^s$, such that
\be
\tilde V_i(x_i) = \tilde V_i(x_i^s) +\half \Omega_i\left(x_i-x_i^s\right)^2 +\ldots,
\ee
with $\Omega_i$ the curvature around the minimum, the exponentials in Eq.\ (\ref{eq:Cgas}) are Gaussians centred at $x_i=x_i^s$ with variance 
\be
\label{eq:sigma}
\sigma_i^2 = \frac{\alpha}{|\cB|}  \frac{\tilde g_i^2}{\Omega_i}.
\ee
The interplay between learning rate, batch size, curvature of the loss function, and the variance of fluctuations  is then easy to see.
 We emphasise that the assumption of a single well-defined minimum is not required in the derivation leading to Eq.\ (\ref{eq:Vi}).

We conclude that the stationary distribution after training depends on the ratio of the learning rate and batch size, and not on these quantities separately. This was observed empirically previously \cite{Smith-1,Smith-2}, but we have demonstrated that it is a direct consequence of stochastic matrix dynamics, when cast in this framework.
Ref.~\cite{DBLP:journals/corr/abs-1710-11029} contains an alternative derivation of the linear scaling rule, using a continuous-time limit of SGD in a weak sense \cite{li2017}, while Ref.~\cite{Sclocchi_2024} contains an application interpreting $\alpha/|\cB|$ as an effective temperature.

The Coulomb gas structure of the stationary distribution has further, universal, consequences. Due to the Coulomb term, eigenvalues repel and the spacing between eigenvalues is non-zero. This effect is captured in the Wigner surmise, which is the distribution for the level spacings $S_i = x_{i+1}-x_i$, see Eqs.\ (\ref{eq:surmise1}, \ref{eq:surmise2}) .
The spectral density, 
\be
\rho(x) = \left\bra \frac{1}{N}\sum_{i=1}^N\delta(x-x_i)\right\ket,
\ee
takes a universal form as well, called the Wigner semicircle, see Eq.\ (\ref{eq:specdens2}).
In the following we explore both universal and non-universal features in two examples.

\section{Applications}
\label{sec:apply}

In this section we apply the general framework presented above to Gaussian restricted Boltzmann machines (RBMs) and to a teacher-student model. In the former, stochasticity is inherently present, due to sampling and batch updates. In the latter, stochasticity is introduced by hand, but in such a way that it models the stochasticity presented above and observed in the RBMs. 

\subsection{Gaussian RBMs}

\begin{figure}[b]
\begin{center}
\includegraphics[width=0.45\textwidth]{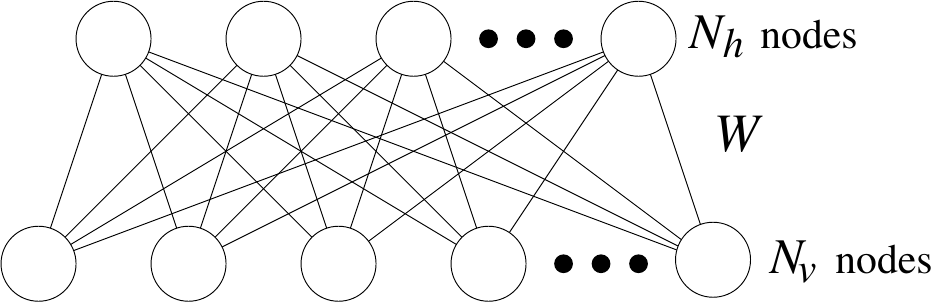}
\end{center}
 \caption{Sketch of a restricted Boltzmann machine with $N_v$ ($N_h$) nodes on the visible (hidden) layer, connected by the $N_v\times N_h$ matrix $W$.
 }
\label{fig:RBM}
\end{figure}

Restricted Boltzmann machines \cite{smolensky,10.1162/089976602760128018} consist of one visible layer (with $N_v$ nodes) and one hidden layer (with $N_h$ nodes), see Fig.\ \ref{fig:RBM}. There are no connections within each layer. The degrees of freedom can be discrete, as in an Ising model, continuous or mixed; Ref.\ \cite{Decelle_2021} is a useful review.
We consider Gaussian RBMs, in which both sets of degrees of freedom are quadratic. The ones on the visible layer are collected in an $N_v$-dimensional vector $\phi$ and on the hidden layer in an $N_h$-dimensional vector $h$. They are coupled bilocally via the $N_v\times N_h$ weight matrix $W$. We follow the notation of our previous work \cite{Aarts:2023uwt}.

The probability distribution and partition function are given by 
\be
p(\phi,h) = \frac{1}{Z} e^{-E(\phi, h)},
\qquad
Z = \int D\phi Dh\, e^{-E(\phi, h)},
\ee
with the energy (or action)
\be
E(\phi, h) = \half \mu^2 \phi^T\phi +\frac{1}{2\sigma_h^2}(h-\eta)^T(h-\eta) - \phi^T W h.
\ee
The induced distribution on the visible layer is Gaussian as well, and reads
\be
p(\phi) = \int Dh\, p(\phi,h) = \frac{1}{Z} \exp\left(
 -\half\phi^T K\phi + J^T\phi \right),
\ee
with the kernel and source
\be
K = \mu^2\id - \sigma_h^2 WW^T, \qqquad 
J = W\eta.
\ee
Here $1/\mu^2$ is the variance on the visible layer, $\sigma_h^2$ is the variance on the hidden layer, and $\eta$ is a bias, which we will put to zero from now on. 
This model has been studied in detail from the perspective of lattice field theory (LFT) in Ref.\ \cite{Aarts:2023uwt}, which explains the appearance of $\mu^2$ as a mass parameter for a scalar field.
Its generative power is linked explicitly to the number of hidden nodes $N_h$ and the mass parameter $\mu^2$, which both act as an ultraviolet regulator, using the terminology familiar from LFT. To match to the general notation of the previous section, we identify $N_v=M$ and $N_h=N$ and take $N\leq M$.

In order to have a well-defined model, with a positive kernel $K=\mu^2\id- \sigma_h^2 WW^T$, the weight matrix $W$ must be initialised in such a way that the eigenvalues of $WW^T$ are bounded between $0$ and $\mu^2/\sigma_h^2$, for any  $0<N/M\leq 1$. Initialisation is discussed in App.\ \ref{app:MP}, in terms of the Marchenko-Pastur distribution.  
Stable initialisation is guaranteed provided the matrix elements of $W$ are drawn from a normal distribution with variance $1/M$ and $\mu^2> 4\sigma_h^2$.

The model can be trained by maximising the log-likelihood (or minimising the Kullback-Leibler divergence), leading to the gradient, 
\be
\label{eq:gradRBM}
\frac{\delta {\cal L}}{\delta W_{ia}} = \sigma_h^2 \sum_j \big( \bra\phi_i\phi_j\ket_{\rm target} - \bra\phi_i\phi_j\ket_{\rm model} \big) W_{ja}.
\ee
Here the first two-point function is evaluated using the target data and the second one is the RBM prediction. 
The dynamics can be analysed semi-analytically \cite{Decelle_2021,Aarts:2023uwt} by performing a singular-value decomposition $W = U \Xi V^T$ in which $U$ and $V$ are orthogonal transformations in $M$ and $N$ dimensions respectively, and $\Xi$ is a rectangular matrix with $N$ singular values $\xi_i$ ($i=1,\ldots, N$) on the diagonal.

The kernel $K$ on the visible layer is diagonalised as
\be
K  = \mu^2\id - \sigma_h^2 WW^T = UD_KU^T,
\ee
where
\begin{align}
& D_K = \nn \\
& \mbox{diag}\big(\underbrace{\mu^2-\sigma_h^2\xi_1^2, \mu^2-\sigma_h^2\xi_2^2, \ldots, \mu^2-\sigma_h^2\xi_{N}^2}_{N}, \underbrace{\mu^2_{ {}_{} }, \ldots, \mu^2}_{M-N}\big).
\end{align}
We are interested in the $N$ eigenvalues $\lambda_i = \mu^2-\sigma_h^2\xi_i^2$. 
It is natural to absorb $\sigma_h^2$ in $x_i= \sigma_h^2\xi_i^2$, such that the eigenvalues read $\lambda_i = \mu^2-x_i$. 
As above, we consider the symmetric matrix $X=W^TW$ (times $\sigma_h^2$) and focus on the eigenvalues $x_i$ of $X$, rather than $\lambda_i$, since this makes the relation with the Coulomb gas direct. Note that positiveness and stability requires that $0\leq x_i < \mu^2$.

\begin{figure}[t]
\begin{center}
\includegraphics[width=0.45\textwidth]{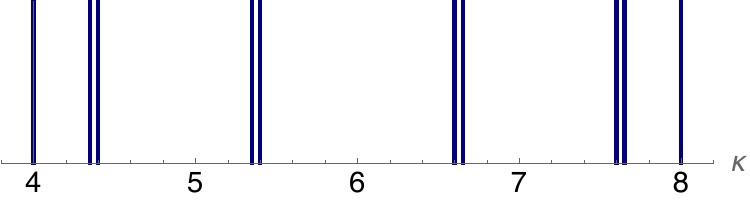}
\end{center}
 \caption{Target spectrum $\kappa_i$ ($i=1,\ldots, 10$): each mode, except the lowest and the highest ones, is doubly degenerate.} 
\label{fig:target-spectrum}
\end{figure}

As the target, we use a simple one-dimensional non-interacting  scalar field theory with the spectrum given by the free dispersion relation, 
\be
\kappa_k = m^2 +p_{{\rm lat}, k}^2 = m^2+2-2\cos\left( \frac{2\pi k}{N_v}\right),
\ee
with $-N_v/2<k\leq N_v/2$. This spectrum is shown in Fig.~\ref{fig:target-spectrum} for the case $N_v=10, m^2=4$. Each mode, except the lowest and the highest ones, is doubly degenerate. This is of interest for the Coulomb gas description derived above, since under stochastic matrix dynamics  eigenvalues repel and exact degeneracy cannot be reproduced.  
Given a target eigenvalue $\kappa_i$ and RBM mass parameter $\mu^2$, the ``exact'' value for $x_i$ is given by $x_i^s = \mu^2-\kappa_i$.

\begin{figure}[h]
\begin{center}
\includegraphics[width=0.45\textwidth]{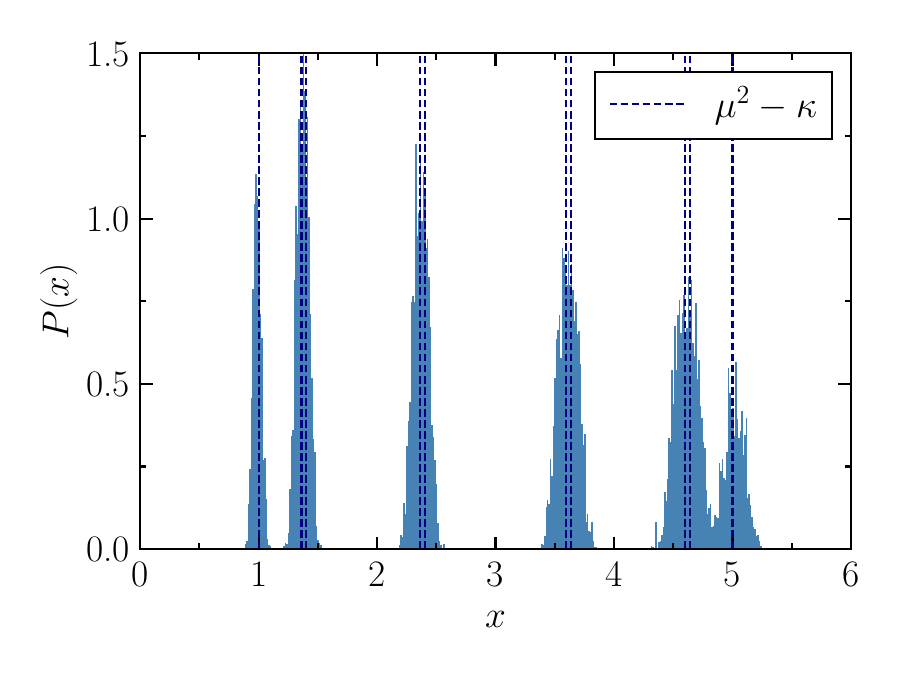}
\end{center}
 \caption{Learnt distributions of eigenvalues $x_i =\mu^2-\lambda_i$. The target eigenvalues $\mu^2-\kappa_i$ are shown with dashed vertical lines. All except the lowest and the highest target eigenvalues are doubly degenerate. 
 }
\label{fig:after-training}
\end{figure}

The numerical training  of the RBM is carried out using persistent contrastive divergence (PCD) with mini-batches, see Ref.~\cite{Aarts:2023uwt} for details. The RBM mass parameter is fixed at $\mu^2=9$ and $N_v=N_h=10$.
 The learning rate and the mini-batch size can be independently varied. For each choice of learning rate and mini-batch size, we have trained 7500 RBMs, to gather statistics. A typical example of the learnt distribution of eigenvalues $x_i$ of $X$ is shown in Fig.\ \ref{fig:after-training}.

\begin{figure*}[t]
\begin{center}
\includegraphics[width=0.48\textwidth]{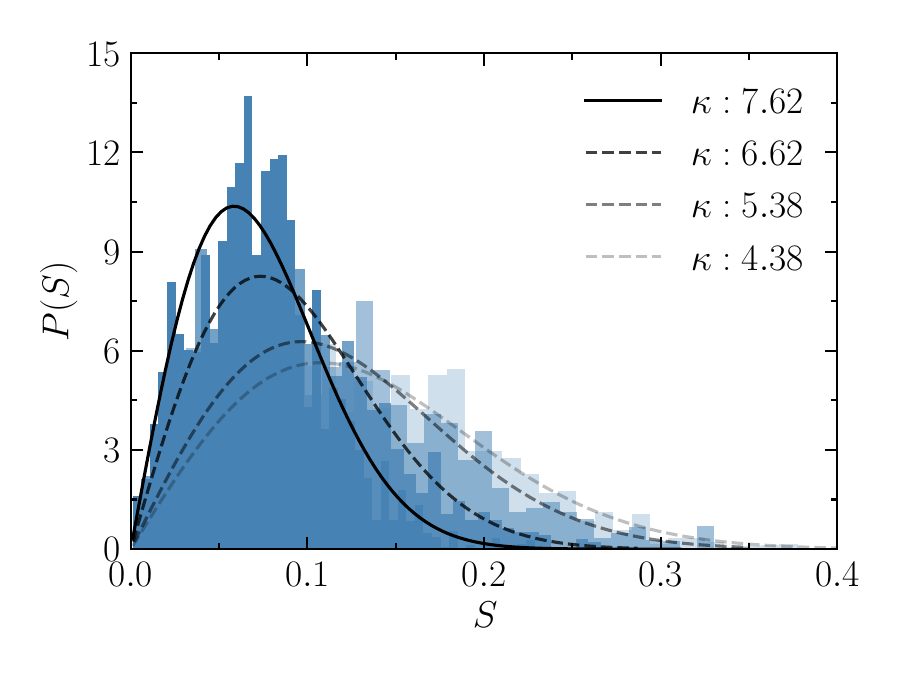}
\includegraphics[width=0.48\textwidth]{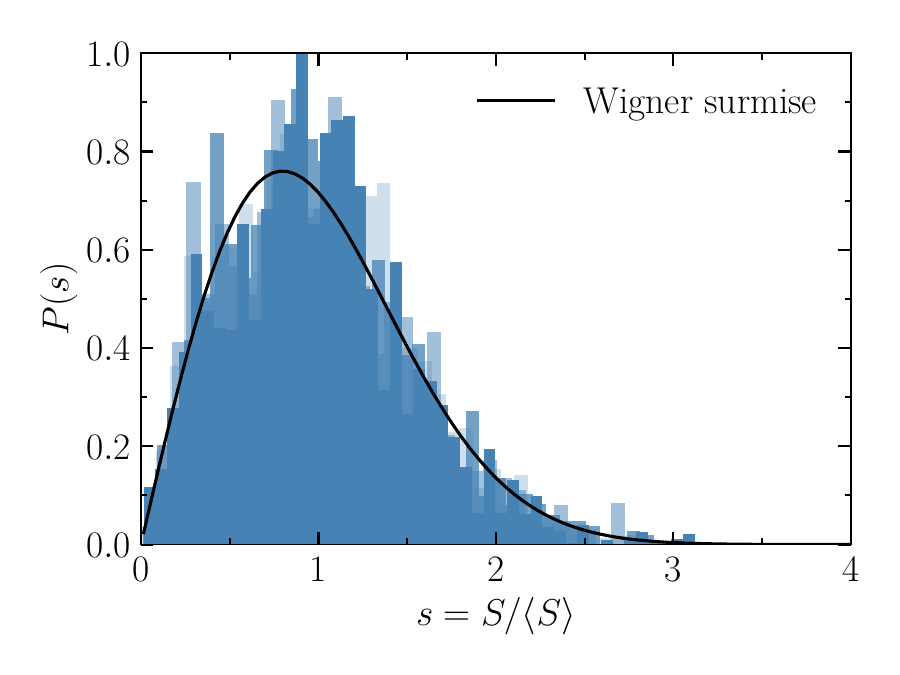}
\end{center}
 \caption{Wigner surmise for the level spacing $S$ (left) and for the rescaled $s=S/\bra S\ket$ (right) for the four doubly-degenerate modes labelled by $\kappa$. The lines on the left are fits with $\bra S\ket = \sqrt{\pi}\sigma$ as a free parameter. The rescaled histograms on the right collapse to the universal curve, $P(s)$.
 }
\label{fig:surmise}
\begin{center}
\includegraphics[width=0.48\textwidth]{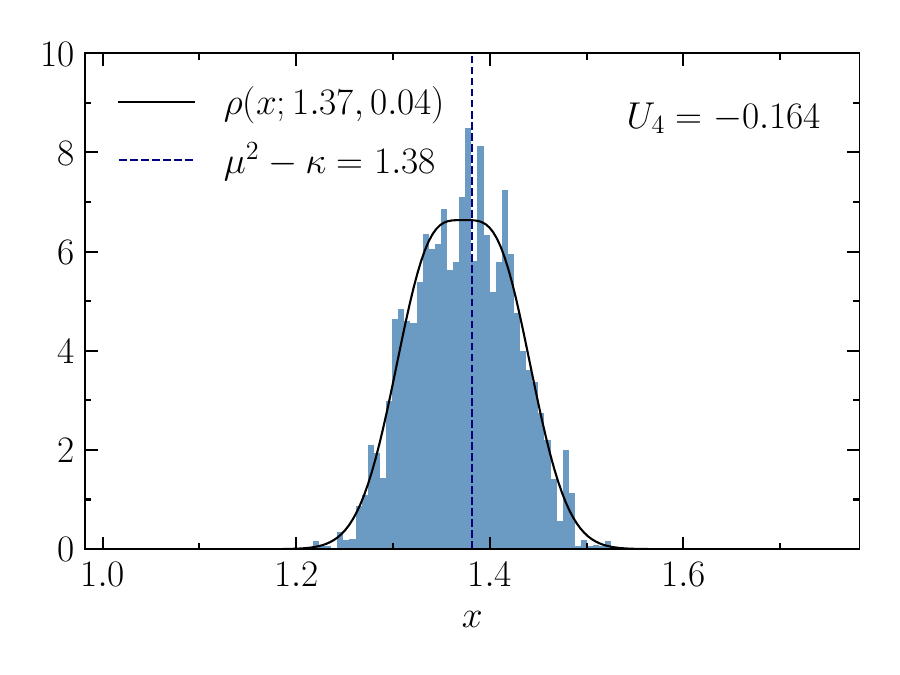}
\includegraphics[width=0.48\textwidth]{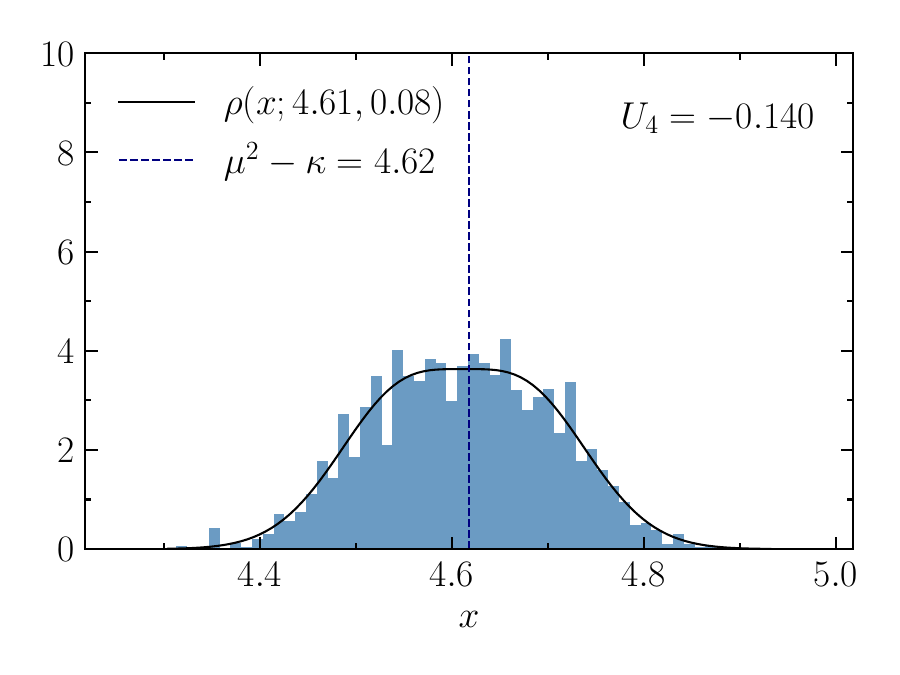}
\end{center}
 \caption{Spectral density for two values of doubly-degenerate target eigenvalue $\kappa=7.62$ (left) and $\kappa= 4.38$ (right), with fit parameters $x_m$ and $\sigma$ indicated in the argument of $\rho(x; x_m,\sigma)$.
 Non-Gaussianity of the distribution is checked by computing the Binder cumulant $U_4$.
 }
\label{fig:spectral-density}
\end{figure*}

\subsection{Universal predictions}

The task is to describe the distributions shown in Fig.\ \ref{fig:after-training} using the concepts from random matrix theory presented above. We start with the universal predictions. Due to the eigenvalue repulsion, degenerate modes cannot be learnt exactly. This can be analysed using the Wigner surmise: for each of the four doubly-degenerate modes we collect data on the level splitting for adjacent eigenvalues, $S_i=x_{i+1}-x_i$. 
Here we employed the fact that the non-degenerate eigenvalues are well-separated. 
The resulting histograms are shown in Fig.~\ref{fig:surmise} (left). These histograms can be fitted with a single-parameter Ansatz -- see Eq.~(\ref{eq:surmise1}) -- in terms of the mean level splitting 
$\bra S_i\ket = \sqrt{\pi}\sigma_i$. Presenting the same data as a function of $s=S_i/\bra S_i\ket$ yields the parameter-free Wigner surmise -- see Eq.~(\ref{eq:surmise2}) -- leading to a universal description of the level spacing. The linear rise at small $s$ reflects the level repulsion.

\begin{figure*}[t]
\begin{center}
\includegraphics[width=0.32\textwidth]{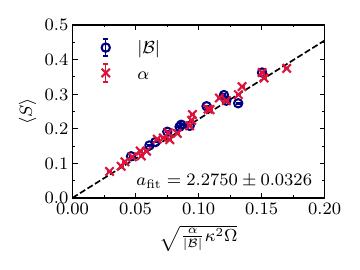}
\includegraphics[width=0.32\textwidth]{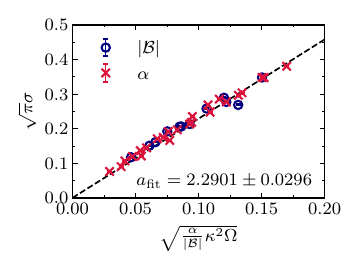}
\includegraphics[width=0.32\textwidth]{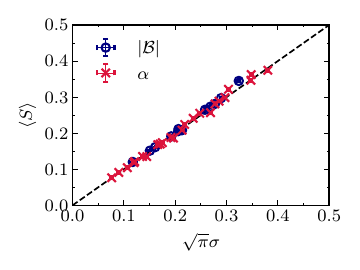}
\end{center}
 \caption{Response of the mean level spacing $\bra S\ket$ (left) and the width parameter of the spectral density $\sqrt{\pi}\sigma$ (middle) to variation of the learning rate $\alpha$ and the batch size $|\cB|$, presented in the combination $\sqrt{(\alpha/|\cB|)\kappa_i^2\Omega_i}$, for 4 doubly-degenerate pairs, identified by target eigenvalues $\kappa_i$. Expected linear relation between $\bra S\ket$ and  $\sqrt{\pi}\sigma$ upon independent variation of $\alpha$ and $|\cB|$ (right).
  }
\label{fig:linear}
\end{figure*}

Next we turn to the spectral density (\ref{eq:specdens1}) for each of the four doubly-degenerate modes. The results for two of these are shown in Fig.~\ref{fig:spectral-density} (note that these are the second-to-lowest and second-to-highest histograms previously shown in Fig.~\ref{fig:after-training}).
The lines are fits to Eq.\ (\ref{eq:specdens2}) with the position $x_{m,i}$ and width $\sigma_i$ as free parameters. It is clear that the histograms are not described by simple Gaussians, but are broader and flatter. Non-Gaussianity can be checked by computing the Binder cumulant $U_4$ of the distribution, which equals (here $\delta x=x-x_m$)
\be
    U_4 \equiv \frac{\left \langle \delta x^4 \right \rangle }{3 \left \langle \delta x^2 \right \rangle ^2} - 1 = -\frac{4}{27} \approx -0.148
\ee
for the Wigner semi-circle (with $N=2$), while it vanishes for a Gaussian distribution.
For completeness, we note here that we have also fitted the distributions corresponding to a single, non-degenerate level (the ones furthest to the left and right in Fig.\ 3) and found these to be described by a Gaussian, as expected.

Importantly, from the general discussion in the previous section it follows that the higher the level of stochasticity (as indicated by $g_i^2$ and depending on learning rate and batch size), the wider the distributions are expected to be. We have hence repeated the exercise above for a range of learning rates, $0.01\leq \alpha\leq 0.1$, and mini-batch sizes, $4\leq |\cB| \leq 64$. 
We will demonstrate below that the potentials $V_i(x_i)$ have a global minimum at $x_i^s=\mu^2 - \kappa_i$. 
The variance in the Coulomb gas $\sigma_i^2$ is then proportional to the ratio of the learning rate and batch size, see Eq.\ (\ref{eq:sigma}), and both the mean level spacing $\bra S\ket$ and the width of the spectral density are expected to scale with $\sqrt{\alpha/|\cB|}$ times a model and data dependent function of $\mu^2$ and $\kappa_i$. 

This is demonstrated in Fig.\ \ref{fig:linear}. We show the response of the mean level spacing $\bra S\ket$ (left) and the width parameter of the spectral density $\sqrt{\pi}\sigma$ (middle) to variation of the learning rate $\alpha$ and the batch size $|\cB|$, as follows,
\be
\bra S\ket, \sqrt{\pi}\sigma = a_{\rm fit} \sqrt{\frac{\alpha}{|\cB|} \kappa_i^2\Omega_i}.
\ee
A derivation of the non-universal factor $\kappa_i^2\Omega_i=(\mu^2-\kappa_i)$, using an analysis of the variance of the gradient of the loss function, can be found in App.\ \ref{app:scaling}.  A linear dependence of $\bra S\ket^2$ and $\sigma^2$ on $\alpha/|\cB|$ is observed, consistent with the derivation in Sec.~\ref{sec:weight}.

Both fits are independent ways to probe the Coulomb gas description of the joint distribution $P(x_i, x_{i+1})$, which depends on one parameter $\sigma_i$ only. Hence in Fig.\ \ref{fig:linear} (right) we show the expected linear relation between  $\bra S\ket$ and $\sqrt{\pi}\sigma_i$, to provide further support for this description.

\begin{figure}[t]
\begin{center}
\includegraphics[width=0.45\textwidth]{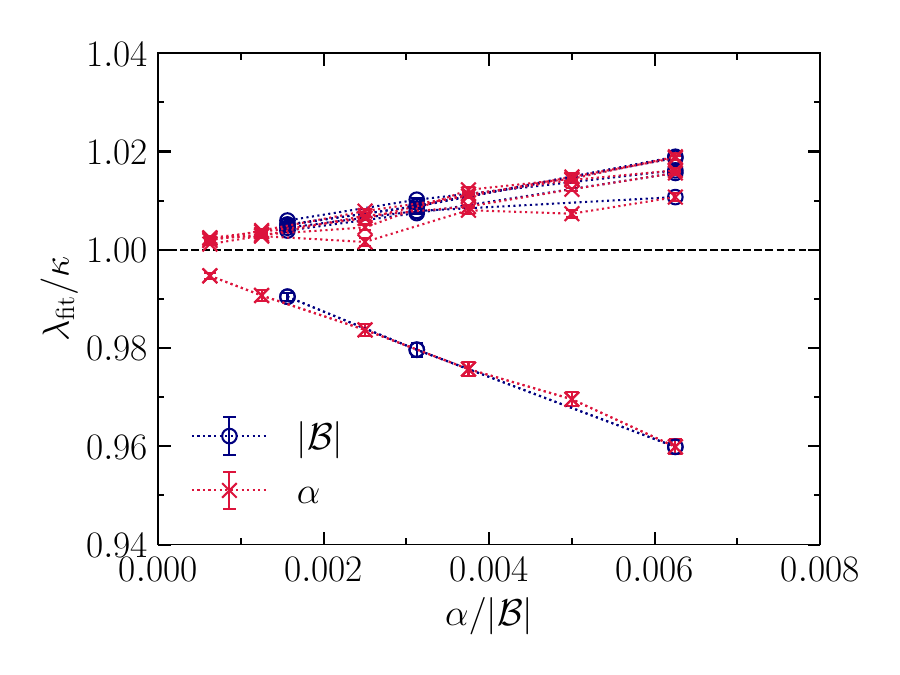}
\end{center}
 \caption{Ratio of the RBM eigenvalues $\lambda_i=\mu^2-x_{m,i}$ and the target eigenvalues $\kappa_i$ as a function of $\alpha/|\cB|$, where $\alpha$ and $|\cB|$ are independently varied,  demonstrating eigenvalue repulsion for non-vanishing stochasticity.
 }
\label{fig:repel}
\end{figure}

The Coulomb term not only breaks the degeneracy of the doubly-degenerate modes, but also leads to a repulsion between all eigenvalues. This can be analysed by determining the mean positions $x_{m,i}$ of the spectral densities as a function of $\alpha/|\cB|$. In Fig.~\ref{fig:repel} we show the ratio of the RBM eigenvalues $\lambda_i=\mu^2-x_{m,i}$ and the target eigenvalues $\kappa_i$ as a function of $\alpha/|\cB|$, where $\alpha$ and $|\cB|$ are independently varied.
We conclude that the spectrum of the target theory will only be learnt exactly in the limit that the stochasticity goes to zero, i.e.\ when $\alpha/|\cB| \to 0$. 
Of course, as is well known, choosing a small learning rate from the beginning of the training is not recommended, since it may lead to overfitting and limit the scope for generalisation.

\subsection{Non-universal dynamics}
  
Next we turn to non-universal, model-dependent features, which are contained in the drift $K_i(x_i)$ and the potential $V_i(x_i)$. In Refs.\ \cite{Decelle_2021, Aarts:2023uwt} it was shown that in the instantaneous eigenbasis the singular values $\xi_i$ of $W$ are subject to the equation
\be
\frac{d}{dt}\xi_i = \sigma_h^2\left( \frac{1}{\kappa_i} -\frac{1}{\mu^2-\sigma_h^2\xi_i^2}\right)\xi_i,
\ee
where it is assumed that a continuous time limit exists.
Writing this as an equation for $x_i=\sigma_h^2\xi_i^2$ (as above) and redefining time as $\tau = 2\sigma_h^2t$ then yields
\be
\frac{d}{d\tau}x_i = \left( \frac{1}{\kappa_i} -\frac{1}{\mu^2-x_i}\right)x_i.
\ee
However, it should be clear from the previous discussion that this equation is not relevant for the evolution of the eigenvalues $x_i$ of $X$ in the realistic case, as it lacks both the Coulomb repulsion and the stochasticity. Instead, we should consider -- see Eq.~(\ref{eq:appx}) --
\be
\frac{d}{d\tau}x_i = K_i(x_i) + \sum_{j\neq i}\frac{g_i^2}{x_i-x_j} + \sqrt{2}g_i\eta_i,
\ee
with a now determined drift
\be
K_i(x_i) = \left( \frac{1}{\kappa_i} -\frac{1}{\mu^2-x_i}\right)x_i.
\ee
Here we stick to continuous time and it is understood that $g_i$ captures the dependence on $\alpha$ and $|\cB|$.

From now on we consider one mode only and drop the index $i$. This allows us to focus on the properties of the drift $K(x)$, without having to consider the Coulomb term, which will be added again later. 
The corresponding Fokker-Planck equation (FPE) for one mode reads
\be
\label{eq:FPE}
\partial_\tau P(x,\tau) = \partial_x\left( g^2\partial_x - K(x)\right)P(x,\tau),
\ee
with the drift
\be 
K(x) = \left( \frac{1}{\kappa} -\frac{1}{\mu^2-x}\right)x.
\ee
A stationary solution exists if the drift can be integrated to yield a potential, using $K(x) = -\partial_x V(x)$, or
\be
V(x) = -\int^x dx'\, K(x') =  -\frac{x^2}{2\kappa} -x-\mu^2\log\left(\mu^2-x\right),
\ee
where we recall that $0 \leq x< \mu^2$, due to positivity and stability requirements. The stationary distribution for one mode then reads
\begin{align}
P_s(x) & = \frac{1}{Z} e^{-V(x)/g^2} \nn \\
&= \frac{1}{Z}\exp\left[ \frac{1}{g^2}\left( \frac{x^2}{2\kappa} + x + \mu^2\log\left(\mu^2-x\right)\right)\right].
\label{eq:Ps}
\end{align}
$Z$ is the normalisation factor, such that 
\be
\int_0^{\mu^2} dx\, P_s(x)=1.
\ee
Note that the distribution is peaked at $x=x_s = \mu^2-\kappa$, where the drift vanishes, $K(x_s)=0$.

To analyse the properties of the time-dependent FPE, we cast the dynamics as a quantum-mechanical bound state problem \cite{Damgaard:1987rr}. We factor out the square root of the stationary distribution 
\be
\label{eq:psi}
P(x,\tau) = \sqrt{P_s(x)} \psi(x,\tau),
\ee
and analyse the time dependence of $\psi(x, \tau)$. 
Taking the time derivative and using Eq.\ (\ref{eq:FPE}) then yields
\begin{align}
& \partial_\tau \psi(x,\tau) = 
\Big( g^2\partial_x^2 - \frac{1}{4g^2}\left[\partial_x V(x)\right]^2 
\nn \\
&\qquad  + \frac{1}{2} \left[\partial_x^2V(x)\right]\Big) \psi(x,\tau) 
 \equiv -2 H_{\rm FP}\psi(x,\tau),
\end{align}
where the so-called Fokker-Planck Hamiltonian can be written in a semi-positive definite form,
\be 
H_{\rm FP} = \half L^\dagger L, 
\ee
with
\be
L^\dagger = -g\partial_x +\frac{1}{2g}\partial_x V(x),
\qquad
L = +g\partial_x +\frac{1}{2g}\partial_x V(x),
\ee
The eigenvalue problem given by $H_{\rm FP}$ takes the form of a bound state problem, with
\be
H_{\rm FP} \psi_n(x) = E_n\psi_n(x).
\ee
The ground state, $\psi_0(x)$, with vanishing energy $E_0=0$, is determined by $L\psi_0(x)=0$, which yields $\psi_0(x) = \sqrt{P_s(x)}$, as it should be, see Eqs.~(\ref{eq:Ps}) and (\ref{eq:psi}). 

The solutions with positive energy determine the time-dependent evolution, or ``learning'' dynamics, and the time-dependent solution of the FPE can be expressed in terms of
\be
\psi(x,\tau) = \psi_0(x) + \sum_{n>0} c_n\psi_n(x) e^{-2E_n\tau},
\ee
where the coefficients $c_n$ are determined by the initial distribution $P(x,0)$, using relation (\ref{eq:psi}).

\begin{figure*}[t]
\begin{center}
\includegraphics[width=0.48\textwidth]{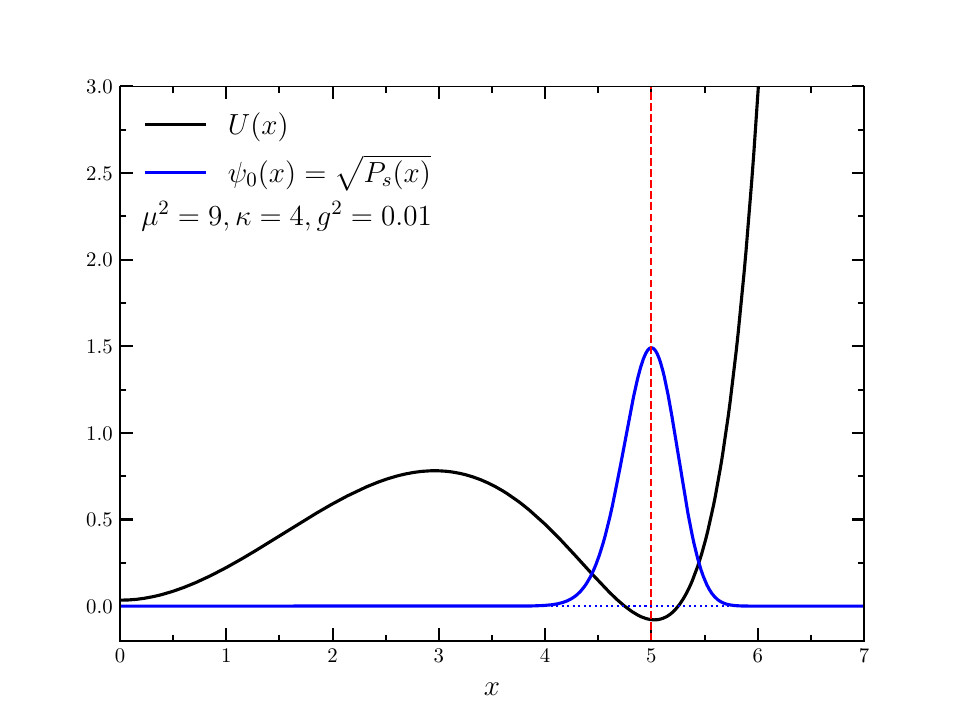}
\includegraphics[width=0.48\textwidth]{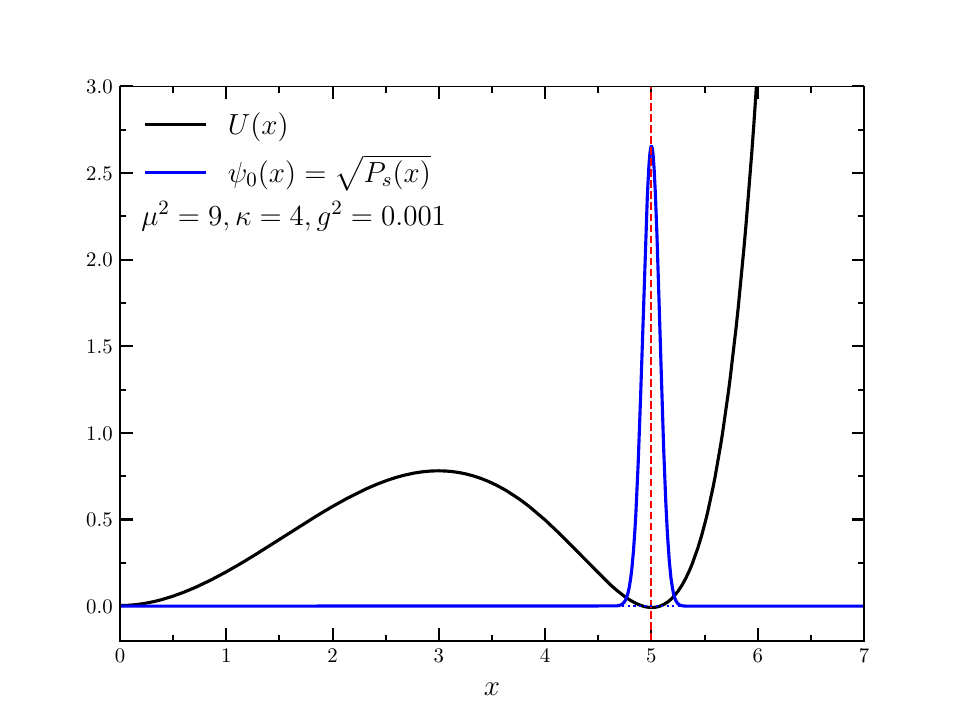}
\end{center}
  \caption{Analysis of the Fokker-Planck Hamiltonian for a single mode $x$, with $\mu^2=9, \kappa=4$, and $g^2=0.01$ (left) and $0.001$ (right). Shown are the FP potential $U(x)$ (black line) and the ground state wave function $\psi_0(x)=\sqrt{P_s(x)}$ (blue line). The vertical dotted line at $x_s=\mu^2-\kappa=5$ indicates the expected position of the peak.  }
    \label{fig:singlemode}
\end{figure*}

To analyse the spectrum of $H_{\rm FP}$ in more detail, we write it as follows,
\be
H_{\rm FP} = -\frac{g^2}{2}\partial_x^2 + U(x), 
\quad
U(x) = \frac{1}{g^2}\left[ U_0(x) + g^2U_1(x)\right],
\ee
where
\be
U_0(x) = \frac{1}{8}\left[\partial_x V(x)\right]^2,
\qquad
U_1(x) = -\frac{1}{4}\partial_x^2V(x),
\ee
are both independent of the noise strength $g^2$. This way of writing reveals the dual role of $1/g^2$: it plays the role of mass, appearing in the kinetic ($\partial_x^2$) and potential ($U_0$) terms in a reciprocal manner. Moreover, given that the noise strength is small, $g^2$ can be treated as a small parameter, such that $U_1$ can be treated as a perturbative correction. We therefore first consider $U_0(x)$. It has a minimum where the drift vanishes, 
\be
\partial_x U_0(x)\big|_{x=x_s} = 0, \qquad  \partial_x V(x)\big|_{x=x_s} = -K(x_s) = 0,
\ee
i.e.\ at the expected position $x_s= \mu^2-\kappa$. The curvature at the minimum is given by
\be
\partial_x^2 U_0(x)\big|_{x=x_s} = \frac{\Omega^2}{4},
\qquad
\Omega \equiv \partial_x^2 V(x) \big|_{x=x_s} = \frac{\mu^2-\kappa}{\kappa^2}.
\ee
Note that $U_0(x_s)=0$ and $U_1(x_s)=-\Omega/4$. 
$U_0(x)$ has a local maximum at $x_{\rm max} = \mu^2-\sqrt{\kappa\mu^2}$, with $U_0(x_{\rm max})>0$, $U_1(x_{\rm max})=0$. We also note that $U(0) =\Omega\kappa/(4\mu^2)>0$.
The potential $U(x)$ is therefore a non-degenerate double well potential on the line segment $0\leq x< \mu^2$, with a global minimum at $x=x_s$. Examples are shown in Fig.\ \ref{fig:singlemode} for two values of $g^2$.

Around $x=x_s$, the FP Hamiltonian can hence be written as
\be
H_{\rm FP} = -\frac{g^2}{2}\partial_x^2 + \frac{1}{2g^2}\left(\frac{\Omega}{2}\right)^2\left(x-x_s\right)^2 -\frac{1}{2}\left(\frac{\Omega}{2}\right),
\ee
making complete the mapping to a harmonic oscillator with mass $1/g^2$, frequency $\Omega/2$, and shifted zero-point energy. Its eigenvalues are therefore
\be
E_n = \frac{\Omega}{2}n, \qquad\quad n=0,1,2,\ldots,
\ee
and convergence to the stationary limit is determined by the smallest non-zero eigenvalue, $2E_1 = \Omega$, as observed before \cite{Aarts:2023uwt}.
Also shown in Fig.\ \ref{fig:singlemode} is the groundstate wave function, $\psi_0(x) = \sqrt{P_s(x)}$. We have a chosen a rather large value of $g^2$ here, such that it has a substantial width. 
Indeed, in the harmonic approximation, $g^2$ does not appear in the spectrum but controls the width of the wave function, and hence the distribution $P(x,t)$. Less stochasticity results in a narrower distribution centred around $x=x_s$, as one would expect.

The overall dynamics combines the evolution in the potential $V_i(x_i)$ for each eigenvalue with the Coulomb repulsion between the eigenvalues, leading to the spectral densities shown in Figs.\ \ref{fig:after-training}, \ref{fig:spectral-density}.

\subsection{Teacher-student model}
\label{sec:ts}

Here we implement the findings in a simple teacher-student model, with an emphasis on how the stochasticity can be modelled, when added by hand. This is of interest, since as we have argued above, the noise does not appear in a standard manner (it is multiplied by the learning rate and not its square root, for instance).

\begin{figure*}[t]
\begin{center}
\includegraphics[width=0.32\textwidth]{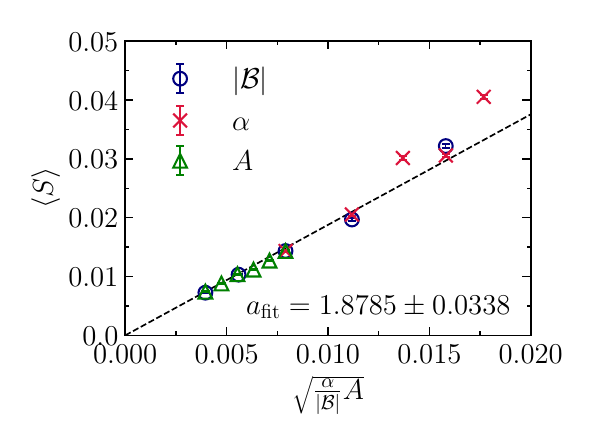}
\includegraphics[width=0.32\textwidth]{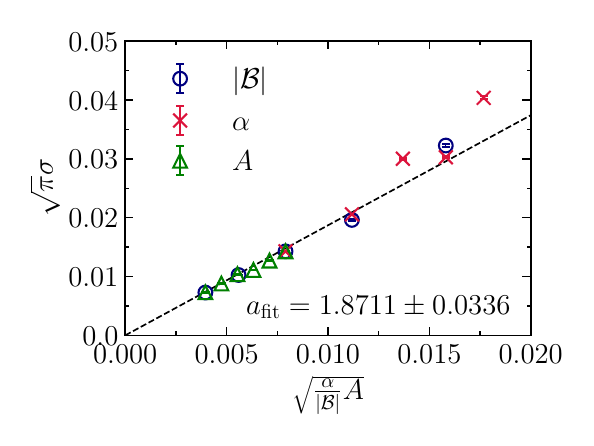}
\includegraphics[width=0.32\textwidth]{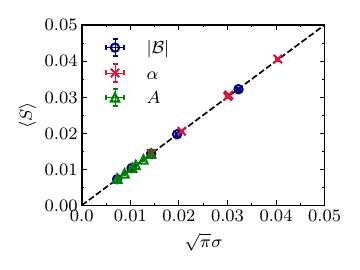}
\end{center}
 \caption{Response of the eigenvalue splitting $\bra S\ket=\bra\delta\lambda\ket$ (left) and the width parameter of the spectral density $\sqrt{\pi}\sigma$ (middle) to independent variations of the learning rate $\alpha$, the batch size $|\cB|$, and the level of stochasticity $A$, presented in the combination $\sqrt{\alpha A/|\cB|}$, in the teacher-student model.
 Expected linear relation between $\bra S\ket$ and  $\sqrt{\pi}\sigma$ upon independent variation of $\alpha$, $|\cB|$, and $A$ (right).
}
 \label{fig:TS}
\end{figure*}

We consider a simple teacher-student model, formulated as
\be
\yv_t = V\xv, \qqquad \yv_s = W\xv,
\ee
where $V, W$ are $N\times N$ symmetric matrices and $\xv \sim \cN(0,1)$ are $N$-dimensional Gaussian random vectors. The teacher matrix $V$ is fixed, while the student matrix $W$ is to be learnt by minimising the loss function,
\be
\cL(W) = \half \left\bra || \yv_t-\yv_s||^2\right\ket,
\ee
where the brackets denote averaging over the input data. In the limit of an infinite amount of input data, $\bra x_ix_j\ket=\delta_{ij}$, and the dynamics is deterministic. Some stochasticity can be introduced by considering batch updates. The amount of stochasticity is, however, not sufficient for the purpose of this discussion, since it is unavoidably suppressed when $W\to V$ as the training converges. We therefore introduce additional randomness by hand, in a manner that is consistent with the discussion in Sec.\ \ref{sec:weight}, see in particular Eq.\ (\ref{eq:Wex}). We consider
\be
W_{ij} \to W_{ij} ' = W_{ij}   - \alpha \frac{\delta \cL}{\delta W_{ij} } + \alpha \sqrt{A_{ij}}  \eta_{ij} ^\cB.
\label{eq:WK}
\ee
Here $\alpha$ is the learning rate, which multiplies both the gradient and the stochastic term, $A_{ij}$ indicates the strength of the noise, and $\eta_{ij}^\cB$ is the stochastic term, averaged over a mini-batch of size $|\cB|$. The latter is introduced as follows: we consider individual updates with noise $\eta_{ij}^b\sim \cN(0,1)$. A mini-batch average is then given by 
\be
\eta_{ij}^\cB = \frac{1}{|\cB|}\sum_{b\in\cB} \eta_{ij}^b,
\ee
such that 
\be
\bra \eta_{ij}^\cB \eta_{kl}^\cB \ket =  \frac{1}{|\cB|^2}\sum_{b,b'}  \bra \eta_{ij}^b\eta_{kl}^{b'}\ket = \frac{1}{|\cB|}\delta_{ik}\delta_{jl}.
\ee
We finally write $\eta_{ij}^\cB = \eta_{ij}/\sqrt{|\cB|}$, with $\eta_{ij} \sim \cN(0,1)$, to separate the factor $1/\sqrt{|\cB|}$. The noise is explicitly symmetrised, such that throughout the evolution $W$ is symmetric.
The final update equation is then
\be
\label{eq:TS}
W_{ij}  \to W_{ij} ' = W_{ij}   - \alpha \frac{\delta \cL}{\delta W_{ij} } + \frac{\alpha}{\sqrt{|\cB|}} \sqrt{A_{ij}}  \eta_{ij},
\ee
with $\eta_{ij} \sim \cN(0,1)$,
which, as stated, is modelled explicitly to be consistent with the arguments given in Sec.~\ref{sec:weight}.

We have applied this model to the case of $N=2$, with the teacher matrix diagonal with eigenvalues $\kappa_{1,2}=\kappa\pm \delta\kappa/2$, and $A_{ij} = A(1+\delta_{ij})$ (see App.\ \ref{app:derivation}). The eigenvalues of $W$ then satisfy
\begin{align}
 \lambda_1\to \lambda_1'  = &\,  \lambda_1 -\alpha(\lambda_1-\kappa_1) +\frac{\alpha^2}{|\cB|}\frac{A}{\lambda_1-\lambda_2} \nn \\
 &+ \frac{\alpha}{\sqrt{|\cB|}}  \sqrt{2A} \eta_1, \\
 \lambda_2\to \lambda_2' = &\,  \lambda_2 -\alpha(\lambda_2-\kappa_2) +\frac{\alpha^2}{|\cB|}\frac{A}{\lambda_2-\lambda_1} \nn \\
& + \frac{\alpha}{\sqrt{|\cB|}}  \sqrt{2A} \eta_2.
\end{align}
We have solved Eq.\ (\ref{eq:TS}) numerically and varied the learning rate $\alpha$, mini-batch size $|\cB|$ and noise strength $A$ independently. We have constructed the distribution for the level spacing and for the spectral density,  and confirmed that in all cases these distributions can be fitted by the Wigner surmise and the Wigner semi-circle (for $N=2$). 

We denote the eigenvalue splitting as $S = |\lambda_1-\lambda_2|$. Taking the difference of the equations above and averaging over the noise, then yields in the stationary limit
\be
\bra S\ket - \delta\kappa -\frac{2\alpha A}{|\cB|}\left\bra\frac{1}{S}\right\ket = 0.
\ee
Since for the Wigner surmise
$\bra 1/S\ket = \pi/(2\bra S\ket)$,
we finally find that the mean level spacing is given by 
\be
\bra S\ket= \half\left( \delta\kappa + \sqrt{\delta\kappa^2 + \frac{4\pi\alpha A}{|\cB|}} \right).
\ee
In the limit of vanishing stochasticy, this becomes $\bra S\ket= \delta\kappa$, as it should be (``perfect learning''), while for the degenerate case this reduces to
\be
\bra S\ket = \sqrt{ \frac{\pi\alpha A}{|\cB|} },
\ee
demonstrating again the linear scaling rule and setting a limit on the precision with which the teacher matrix can be learnt, for finite learning rate, noise strength, and batch size.
In Fig.\ \ref{fig:TS} we show the mean level spacing as a function of $\sqrt{\alpha A/|\cB|}$, where, as already mentioned, $\alpha$, $|\cB|$ and $A$ are varied independently, and $\delta\kappa=0$. The fitted slope, $a_{\rm fit}\approx 1.87$, is close to the analytical prediction $\sqrt{\pi}\approx 1.77$, where the latter is valid when finite learning rate corrections are disregarded (see App.~\ref{app:scaling}).

\section{Summary and outlook}
\label{sec:con}

We have considered stochastic weight matrix dynamics in generic learning algorithms and argued that it can be described in the framework of Dyson Brownian motion. This results in universal features dictated by random matrix theory, including eigenvalue repulsion, quantified by the Wigner surmise. The level of stochasticity is determined by the ratio of the learning rate $\alpha$ and the mini-batch size $|\cB|$ times a non-universal function which encodes details of the loss function and the architecture. The linear scaling rule, i.e. the dependence on $\alpha/|\cB|$, arises naturally within this framework. One implication is that there is no straightforward limit in which, say, stochastic gradient descent reduces to a stochastic differential equation in continuous time, since the level of stochasticity decreases with decreasing learning rate or increasing mini-batch size. While the inherent stochasticity for finite $\alpha/|\cB|$ sets a fundamental limit on the accuracy of learning, it also prevents overfitting and hence helps in decreasing the generalisation error.

We have tested our universal and non-universal predictions in two simple models: a linear teacher-student model and the Gaussian restricted Boltzmann machine. In the latter the non-universal aspects of the dynamics of the eigenvalues can be nicely described as a quantum-mechanical bound state problem in a non-degenerate double well potential, defined on a finite interval. The universal dependence on $\alpha/|\cB|$ was confirmed in both models.

Our results are rather general and can be applied to any model in which weight matrices are updated with a stochastic optimisation algorithm.
The choice of architecture and loss function determine the non-universal aspects in the Coulomb potential, as well as the level of stochasticity. In some cases the resulting stochasticity is so small that is hardly noticeable; this was the case in the linear teacher-student model and for that reason we added noise by hand. The Gaussian RBM is stochastic by itself, due to the need to sample, and hence the noise inherent in the model is strong enough to observe the effects without any further addition.
Moving forward, it is important to test these ideas in more complicated architectures, including neural networks and transformers. To be able to observe universal features, it may be required to use spectral unfolding, see e.g.\ Refs.~\cite{Verbaarschot:2000dy,AbulMagd:2014,Levi:2023qwg}. The coupling between different weight matrices across layers as well as the use of adaptive learning rates, such as Adam \cite{kingma:2017}, may also influence the manner in which universal features appear. Work along these directions is currently in progress.

\vspace*{0.2cm} 

\noindent
{\bf Acknowledgements} --  
The content of this paper was first presented at the ECT* workshop {\it Machine Learning and the Renormalisation Group} in May 2024. We thank the participants of this meeting, in particular Ouraman Hajizadeh and Semon Rezchikov, for discussion, and ECT* for support. 
We thank Prem Kumar and Neil Talwar for discussion on matrix models.
GA and BL are supported by STFC Consolidated Grant ST/T000813/1. 
BL is further supported by the UKRI EPSRC ExCALIBUR ExaTEPP project EP/X017168/1.
CP is supported by the UKRI AIMLAC CDT EP/S023992/1.

\noindent
{\bf Research Data and Code Access} --
The code and data used for this manuscript are available from Ref.\ \cite{zenodo}.

\noindent
{\bf Open Access Statement} -- For the purpose of open access, the authors have applied a Creative Commons Attribution (CC BY) licence to any Author Accepted Manuscript version arising.


\appendix

\section{Dyson Brownian motion and the stochastic Coulomb gas}
\label{app:dyson}
  
In this Appendix we summarise the main concepts of  Dyson Brownian motion \cite{Dyson-4}, see in particular the text book \cite{Meh2004} (Ch.\ 7). For simplicity we use continuous-time notation, but when in doubt, expressions should be understood in the sense of It\^o calculus.

Consider a symmetric $N\times N$ matrix $X$, whose matrix elements are updated according to
\be
\frac{dX_{ij}}{dt} = K_{ij}(X) + \sqrt{A_{ij}}\eta_{ij},
\ee
where $K_{ij}$ is the drift term, $A_{ij}$ encodes the strength of the noise, and $\eta_{ij}\sim \cN(0,1)$, all of which are symmetric.
 Its eigenvalues $x_i$ then evolve according to
\begin{align}
\frac{dx_i}{dt} &= K_i(x_i) + \sum_{j\neq i}\frac{g_i^2}{x_i-x_j} + \sqrt{2}g_i\eta_i 
\nn\\
&\equiv K_i^{\rm (eff)}(x_i) + \sqrt{2}g_i\eta_i,
\label{eq:appx}
\end{align}
where $K_i(x_i)$ is the drift acting on the eigenvalues, the second term is the induced Coulomb term, and $\eta_i\sim \cN(0,1)$. This result can be derived using second-order perturbation theory with a discrete time step $\delta t$, making the standard assumption that the drift and the noise correlator $\bra\eta_{ij}\eta_{kl}\ket$ scale as $\delta t$ \cite{Meh2004}.
The noise strength satisfies $A_{ij} = A(1+\delta_{ij})$ and we denote the diagonal element as $\sqrt{A_{ii}}=\sqrt{2}g_i$. Strictly speaking the index on $g_i$ should be dropped, but we keep it for future convenience. 
A derivation in the context of the teacher-student model is given in App.~\ref{app:derivation}.

The corresponding Fokker-Planck equation (FPE) for the distribution $P(\{x_i\},t)$ reads
\be
\partial_t P(\{x_i\},t) 
= \sum_{i=1}^N \partial_{x_i} \left[  \left( g_i^2\partial_{x_i} - K_i^{\rm (eff)}(x_i)\right)\right] P(\{x_i\},t).
\ee
Provided that the drift can be derived from a potential,
\be
\label{eq:Ki}
K_i(x_i) = -\frac{dV_i(x_i)}{dx_i},
\ee
the FPE has the stationary solution
\be
\label{eq:PV}
P_s(\{x_i\}) =  \frac{1}{Z}\prod_{i<j} \left|x_i-x_j\right| e^{-\sum_i V_i(x_i)/g_i^2},
\ee
with 
\be
Z = \int dx_1 \ldots dx_N\, P_s(\{x_i\}).
\ee
This distribution is known as the Coulomb gas. The Coulomb repulsion term can also be seen as a Jacobian 
arising when making the transition from integrating over all matrix elements to just the eigenvalues. 
In the case that the potentials $V_i(x_i)/g_i^2$ are identical quadratics, the Coulomb gas describes the eigenvalues of the Gaussian orthogonal ensemble.  

The presentation above assumes that the potential in the Coulomb gas is separable, see Eqs.~(\ref{eq:Ki}, \ref{eq:PV}). If this is not possible, e.g.\ due to eigenvalues mixing non-linearly in the drift terms, a more general potential appears, as
\be
 \sum_i V_i(x_i)/g_i^2 \to \overline{V}(x_1, x_2, \ldots, x_N),
 \ee
 with
 \be
 K_i = -g_i^2\frac{\partial\overline{V}(x_1, x_2, \ldots, x_N)}{\partial x_i}.
 \ee
 While this will complicate the analysis and lead to a richer loss function landscape, the implications are expected to remain.

We specialise to the case with $N=2$ eigenvalues and assume that the dynamics can be linearised around degenerate minima $x_1^s=x_2^s=\kappa$, such that the drift $K_i$ is proportional to $(\kappa-x_i)$. The partition function is then
\be
Z = \frac{1}{N_0}\int dx_1dx_2\, |x_1-x_2| \, e^{-V(x_1,x_2)},
\ee
with 
\be
V(x_1, x_2) = \frac{1}{2\sigma^2}\left[ (x_1-\kappa)^2 +  (x_2-\kappa)^2\right].
\ee
The variance $\sigma^2$ is chosen to be identical for the two degenerate modes. We assume that the distribution is sufficiently peaked around $\kappa$, such that the integral boundaries can be taken as $\pm\infty$. We hence consider 
\be
Z = \frac{1}{N_0}\int dx_1dx_2\, |x_1-x_2| \, e^{- \left(x_1^2 + x_2^2\right)/(2\sigma^2)},
\ee
with the normalisation constant $N_0=4\sqrt{\pi}\sigma^3$.

The Wigner surmise signifies the level spacing $S=x_1-x_2$. Let us change variables to $x_{1,2}=x\pm S/2$, such that
\be
Z =  \int_0^\infty dS\, P(S),
\ee
with
\be
\label{eq:surmise1}
P(S) =  \frac{S}{2\sigma^2} e^{-S^2/(4\sigma^2)}.
\ee
The mean level spacing is
\be
\bra S\ket =  \int_0^\infty dS\, S P(S) = \sqrt{\pi}\sigma.
\ee
In terms of $s=S/\bra S\ket$ the surmise is parameter-free,
\be
\label{eq:surmise2}
P(s) = \frac{\pi}{2} s e^{-\pi s^2/4}.
\ee

The spectral density is defined, for arbitrary $N$, as 
\be
\label{eq:specdens1}
\rho(x) = \left\bra \frac{1}{N}\sum_{i=1}^N\delta(x-x_i)\right\ket.
\ee
For $N=2$, it is easily evaluated as
\be
\label{eq:specdens2}
\rho(x) = \frac{e^{-x^2/(2\sigma^2)}}{4\sqrt{\pi}\sigma} \left[ 2e^{-x^2/(2\sigma^2)} +\sqrt{2\pi} \frac{x}{\sigma}\mbox{Erf}\left(\frac{x}{\sqrt{2}\sigma} \right)\right].
\ee
It is flatter and broader than a simple Gaussian.

\section{Weight matrix initialisation}
\label{app:MP}

Consider the $M\times N$ random matrix $W$ at initialisation, with elements $W\sim \cN(0,\sigma_{\rm in}^2)$. Without loss of generality we take $N\leq M$; if this is not the case, simply exchange $W$ and $W^T$. The symmetric $N\times N$ matrix $X = W^TW$ has $N$ non-zero eigenvalues $x_i=\xi_i^2$ ($i=1,\ldots, N$), where $\xi_i$ are the singular values of $W$.
The distribution of eigenvalues $x$ of $X$ is given by the Marchenko-Pastur distribution, 
\be
P_{\rm MP}(x) = \frac{1}{2\pi\sigma_{\rm in}^2M r x}\sqrt{(x_+-x)(x-x_-)},
\ee
where $r=N/M\leq 1$ and $x_\pm = M\sigma_{\rm in}^2\left(1\pm\sqrt{r}\right)^2$, $x_-<x<x_+$.
It is desirable for the spectrum of $X$ to only depend on $r$ and not separately on $M$ and $N$, such that one can take $M,N\to \infty$ at fixed $r$. 
There is some freedom to select $\sigma_{\rm in}^2$, e.g.\  $\sigma_{\rm in}^2 = 1/M, 1/N$, or $1/\sqrt{MN}$. 
By choosing $\sigma_{\rm in}^2=1/M$, the spectrum of $X$ is bounded for all values of $0<r\leq 1$, which can be advantageous, in particular in the RBM case. The initial distribution then reads
\be
P_{\rm MP}(x) = \frac{1}{2\pi r x}\sqrt{(x_+-x)(x-x_-)}, 
\ee
with $x_\pm = \left(1\pm\sqrt{r}\right)^2$, $0\leq x_-\leq x\leq x_+\leq 4$.

\section{Non-universal part of the linear scaling relation}
\label{app:scaling}

In Sec.\ \ref{sec:weight}, we demonstrated that the Coulomb gas potential $V_i(x_i)/g_i^2$ is the product of a universal and a non-universal factor, see Eqs.~(\ref{eq:Vi}, \ref{eq:sigma}).
Here we consider the non-universal factor in the case of the restricted Boltzmann machine and sketch a derivation of its parameter dependence.  

We are interested in the variance of the gradient of the loss function, $\Var\left(\delta W\right)$ or $\Var\left(\delta X\right)$. We write the gradient in the form, see Eq.~(\ref{eq:gradRBM}) and putting $\sigma_h^2=1$,
\be
\frac{\delta {\cal L}}{\delta W}W^T = \left( C^{\rm target} -C^{\rm RBM}\right) (WW^T).
\ee
The two-point functions on the RHS indicate the target data,
\be
C^{\rm target} = K_{\rm target}^{-1},
\ee
and the RBM model,
\be
C^{\rm RBM} = \phi^{\rm RBM} \phi^{\rm RBM},
\ee
with expectation value
\be
\label{eq:C}
\left\bra C^{\rm RBM} \right\ket = K^{-1}, \qqquad K = \mu^2\id-WW^T.
\ee
Finally, using the singular-value decomposition $W=U\Xi V^T$, 
\be 
WW^T = U \Xi\Xi^T U^T, 
\ee
where $\Xi\Xi^T$ contains the squares of the singular values $x_i=\xi_i^2$ on the diagonal.  
All quantities above are matrix-valued with indices on the visible layer, which we suppress.

To determine the variance, it is important to note that only $C^{\rm RBM}$ is fluctuating during the mini-batch and contrastive divergence updates. The variance is therefore
\begin{align}
&\Var\left(\frac{\delta {\cal L}}{\delta W}W^T\right) = 
\nn\\
&
\left( \left\bra C^{\rm RBM}C^{\rm RBM}\right\ket 
- \left\bra C^{\rm RBM}\right\ket^2 \right)
 (WW^T)(WW^T).
\end{align}
In the Gaussian RBM only the second moment is non-trivial, see Eq.~(\ref{eq:C}). Hence
\be
\left\bra C^{\rm RBM}C^{\rm RBM}\right\ket - \left\bra C^{\rm RBM}\right\ket^2  = 2 K^{-1} K^{-1}.
\ee
Going to the eigenbasis for one mode, in which 
\be
K^{-1} \to \frac{1}{\mu^2-x}, \qqquad WW^T\to x,
\ee
then yields
\be
\Var\left(\frac{\delta {\cal L}}{\delta W}W^T\right) \sim  \frac{2x^2}{(\mu^2-x)^2}.
\ee
After training, $x=\mu^2-\kappa +\delta x$. At leading order in $\delta x$ we find therefore
\be
\Var\left(\frac{\delta {\cal L}}{\delta W}W^T\right) \sim  \frac{2\left(\mu^2-\kappa\right)^2}{\kappa^2}.
\ee
In the main text we noted that the curvature of the Coulomb gas potential around its minimum is given by $\Omega_i=(\mu^2-\kappa_i)/\kappa_i^2$. We therefore write the non-universal function representing the variance as
\be
\tilde g_i^2 \sim \kappa_i^2 \Omega_i^2,
\ee
or, see Eq.~(\ref{eq:sigma}),
\be
\sigma_i^2 = \frac{\alpha}{|\cB|}\kappa_i^2\Omega_i.
\ee
By analysing the scaling of the mean level splitting $\bra S\ket$ with $\alpha/|\cB|$ for different $\kappa_i$ values, we have confirmed this parametric dependence on $\mu^2$ and~$\kappa$.

As a side note, we also include here the (well-known) statement that stochastic equations suffer from finite-discretisation effects, which should be considered when comparing with analytically derived expressions. Consider the simple update for one degree of freedom, which is modelled according to the equations considered in the main paper,
\be
x_{n+1} = x_n -\alpha \omega x_n + \sqrt{2\alpha\gamma}\eta_n,
\ee
where 
 $\gamma = \alpha A/|\cB|$ and $\bra\eta_n\eta_m\ket= \delta_{nm}$.
It is solved by (taking $x_0=0$) 
\be
x_n = \sqrt{2\alpha\gamma}\sum_{i=0}^{n-1}(1-\alpha\omega)^{n-1-i}\eta_i.
\ee
Assuming that $\alpha\omega< 1$ and taking $n$ large, one finds 
\be
\lim_{n\to \infty} \bra x_n^2\ket = \frac{\alpha A}{|\cB|} \frac{1}{\omega} \frac{1}{1-\alpha\omega/2}.
\ee
This illustrates that the variance depends on $\alpha A/|\cB|$, as we have emphasised in the main text, but also that there are finite learning rate corrections present, as expected.

\section{Derivation of the eigenvalue equation}
\label{app:derivation}

In this appendix we add a brief derivation of the eigenvalue equation including the Coulomb term, starting with the matrix update, in the context of the teacher-student model. This follows closely text book derivations \cite{Meh2004}, but may be useful to those not familiar with this topic. 

We consider a $2\times 2$ symmetric matrix $W$, updated according to 
\be
W_{ij}  \to W_{ij} ' = W_{ij}   - \alpha \frac{\delta \cL}{\delta W_{ij} } + \frac{\alpha}{\sqrt{|\cB|}} \sqrt{A_{ij}}  \eta_{ij} , 
\ee
where $\alpha$ is the learning rate and $|\cB|$ the batch size. 
Both the noise $\eta_{ij} \sim \cN(0,1)$ and strength $A_{ij}$ are symmetric matrices. The eigenvalues of $W$ are denoted as $\lambda_{1,2}$ and the teacher values are $\kappa_{1,2}$. 

We assume that at the current step $W$ is diagonal (or has been diagonalised) and consider one update. The drift is then $-\alpha(\lambda_{1,2}-\kappa_{1,2})$ and the updated matrix reads 
\begin{widetext}
\be
W' =\begin{pmatrix}
\lambda_1 - \alpha(\lambda_1-\kappa_1) + (\alpha\sqrt{|\cB|})\sqrt{A_{11}}\eta_{11} & (\alpha/\sqrt{|\cB|})\sqrt{A_{12}}\eta_{12} \\
(\alpha/\sqrt{|\cB|})\sqrt{A_{12}}\eta_{12} &   \lambda_2 - \alpha(\lambda_2-\kappa_2) + (\alpha/\sqrt{|\cB|})\sqrt{A_{22}}\eta_{22} 
\end{pmatrix}.
\ee
\end{widetext}
To compute the eigenvalues of $W'$, we follow the standard power counting rules and treat the drift as $\cO(\eps)$ and the noise terms as $\cO(\sqrt{\eps})$. Expanding to $\cO(\eps)$ then yields the eigenvalues of $W'$,
\begin{align}
& \lambda_1' = \lambda_1 -\alpha(\lambda_1-\kappa_1) +\frac{\alpha^2}{|\cB|}\frac{A_{12}\eta_{12}^2}{\lambda_1-\lambda_2} + \frac{\alpha}{\sqrt{|\cB|}}  \sqrt{A_{11}} \eta_{11}, 
 \\
& \lambda_2' = \lambda_2 -\alpha(\lambda_2-\kappa_2) +\frac{\alpha^2}{|\cB|}\frac{A_{12}\eta_{12}^2}{\lambda_2-\lambda_1} + \frac{\alpha}{\sqrt{|\cB|}}  \sqrt{A_{22}} \eta_{22}.
\end{align}
Taking a noise average in the Coulomb term, i.e., replacing $\eta_{12}^2$ by $1$, denoting $\eta_{11}=\eta_1, \eta_{22}=\eta_2$, and using that
$A_{ij}=A(1+\delta_{ij})$ (or  $A_{11}=A_{22}=2A, A_{12}=A_{21}=A$)~\cite{Meh2004},
then yields
\begin{align}
& \lambda_1' = \lambda_1 -\alpha(\lambda_1-\kappa_1) +\frac{\alpha^2}{|\cB|}\frac{A}{\lambda_1-\lambda_2} + \frac{\alpha}{\sqrt{|\cB|}}  \sqrt{2A} \eta_1, 
 \\
& \lambda_2' = \lambda_2 -\alpha(\lambda_2-\kappa_2) +\frac{\alpha^2}{|\cB|}\frac{A}{\lambda_2-\lambda_1} + \frac{\alpha}{\sqrt{|\cB|}}  \sqrt{2A} \eta_2.
\end{align}
This is the standard equation for the eigenvalues, including the Coulomb and stochastic terms. As mentioned, a detailed discussed of the various steps and an extension to $N\times N$ matrices can be found in the text book~\cite{Meh2004}.

\providecommand{\href}[2]{#2}\begingroup\raggedright\endgroup


\begin{thebibliography}{10}


\bibitem{Carleo_2019}
G.~Carleo, I.~Cirac, K.~Cranmer, L.~Daudet, M.~Schuld, N.~Tishby et~al.,
  \emph{Machine learning and the physical sciences},
  \href{https://doi.org/10.1103/revmodphys.91.045002}{\emph{Reviews of Modern
  Physics} {\bfseries 91} (2019) 045002}
  [\href{https://arxiv.org/abs/1903.10563}{{\ttfamily 1903.10563}}].

\bibitem{Dyson-4}
F.J.~Dyson, \emph{{A Brownian-Motion Model for the Eigenvalues of a Random
  Matrix}}, \href{https://doi.org/10.1063/1.1703862}{\emph{J. Math. Phys.}
  {\bfseries 3} (1962) 1191}.

\bibitem{Wigner-1}
E.P.~Wigner, \emph{Characteristic vectors of bordered matrices with infinite
  dimensions}, {\emph{Annals of Mathematics} {\bfseries 62} (1955) 548}.

\bibitem{Wigner-2}
E.P.~Wigner, \emph{{Conference on Neutron Physics by Time-of-Flight}},  p.~67,
  1956.

\bibitem{Dyson-1}
F.J.~Dyson, \emph{{Statistical theory of the energy levels of complex systems.
  I}}, \href{https://doi.org/10.1063/1.1703773}{\emph{J. Math. Phys.}
  {\bfseries 3} (1962) 140}.

\bibitem{Dyson-2}
F.J.~Dyson, \emph{{Statistical theory of the energy levels of complex systems.
  II}}, \href{https://doi.org/10.1063/1.1703774}{\emph{J. of Math. Phys.}
  {\bfseries 3} (1962) 157}.

\bibitem{Dyson-3}
F.J.~Dyson, \emph{{Statistical Theory of the Energy Levels of Complex Systems.
  III}}, \href{https://doi.org/10.1063/1.1703775}{\emph{J. Math. Phys.}
  {\bfseries 3} (1962) 166}.

\bibitem{Meh2004}
M.L.~Mehta, \emph{{Random Matrices}}, {Academic Press, New York}, 3rd~ed.
  (2004).

\bibitem{Goyal-1}
P.~Goyal, P.~Doll{\'{a}}r, R.B.~Girshick, P.~Noordhuis, L.~Wesolowski,
  A.~Kyrola et~al., \emph{{Accurate, Large Minibatch {SGD:} Training ImageNet
  in 1 Hour}}, (2017)
  [\href{https://arxiv.org/abs/1706.02677}{{\ttfamily 1706.02677}}].

\bibitem{Smith-1}
S.L.~Smith and Q.V.~Le, \emph{{A Bayesian Perspective on Generalization and
  Stochastic Gradient Descent}}, (2017)
   [\href{https://arxiv.org/abs/1710.06451}{{\ttfamily 1710.06451}}].

\bibitem{Smith-2}
S.L.~Smith, P.~Kindermans and Q.V.~Le, \emph{{Don't Decay the Learning Rate,
  Increase the Batch Size}},  (2017)
  [\href{https://arxiv.org/abs/1711.00489}{{\ttfamily 1711.00489}}].

\bibitem{Smith-3}
S.L.~Smith, D.~Duckworth, Q.V.~Le and J.~Sohl{-}Dickstein, \emph{{Stochastic
  natural gradient descent draws posterior samples in function space}},
   (2018) 
  [\href{https://arxiv.org/abs/1806.09597}{{\ttfamily 1806.09597}}].

\bibitem{DBLP:journals/corr/abs-1710-11029}
P.~Chaudhari and S.~Soatto, \emph{{Stochastic gradient descent performs
  variational inference, converges to limit cycles for deep networks}},
  \href{https://arxiv.org/abs/1710.11029}{{\ttfamily 1710.11029}}.

\bibitem{mandt2015}
S.~Mandt, M.D.~Hoffman and D.M.~Blei, \emph{Continuous-time limit of stochastic
  gradient descent revisited},  in \emph{8th NIPS Workshop on Optimization for
  Machine Learning}, 2015.

\bibitem{li2017}
Q.~Li, C.~Tai and W.~E, \emph{Stochastic modified equations and adaptive
  stochastic gradient algorithms},  in \emph{Proceedings of the 34th
  International Conference on Machine Learning}, vol.~70, pp.~2101--2110, 2017
  [\href{https://arxiv.org/abs/1511.06251}{{\ttfamily 1511.06251}}].

\bibitem{yaida2018}
S.~Yaida, \emph{Fluctuation-dissipation relations for stochastic gradient
  descent},  in \emph{International Conference on Learning Representations},
  2019 [\href{https://arxiv.org/abs/1810.00004}{{\ttfamily 1810.00004}}].

\bibitem{Martin-2019}
C.H.~Martin and M.W.~Mahoney, \emph{{Traditional and Heavy-Tailed Self
  Regularization in Neural Network Models}},  (2019)  [\href{https://arxiv.org/abs/1901.08276}{{\ttfamily
  1901.08276}}].

\bibitem{Baskerville}
N.P.~Baskerville, D.~Granziol and J.P.~Keating, \emph{{Applicability of Random
  Matrix Theory in Deep Learning}}, (2021)  [\href{https://arxiv.org/abs/2102.06740}{{\ttfamily 2102.06740}}].

\bibitem{Levi:2023qwg}
N.~Levi and Y.~Oz, \emph{{The Universal Statistical Structure and Scaling Laws
  of Chaos and Turbulence}},
  \href{https://arxiv.org/abs/2311.01358}{{\ttfamily 2311.01358}}.

\bibitem{Sclocchi_2024}
A.~Sclocchi and M.~Wyart, \emph{{On the different regimes of stochastic
  gradient descent}},
  \href{https://doi.org/10.1073/pnas.2316301121}{\emph{Proceedings of the
  National Academy of Sciences} {\bfseries 121} (2024) }
  [\href{https://arxiv.org/abs/2309.10688}{{\ttfamily 2309.10688}}].


\bibitem{smolensky}
P.~Smolensky, \emph{Chapter 6: Information processing in dynamical systems:
  Foundations of harmony theory},  in \emph{Parallel Distributed Processing:
  Volume 1}, D.~Rumelhart and J.~McLelland, eds., pp.~194--281, MIT Press,
  1986.

\bibitem{10.1162/089976602760128018}
G.E.~Hinton, \emph{{Training Products of Experts by Minimizing Contrastive
  Divergence}}, \href{https://doi.org/10.1162/089976602760128018}{\emph{Neural
  Computation} {\bfseries 14} (2002) 1771}.

\bibitem{Decelle_2021}
A.~Decelle and C.~Furtlehner, \emph{{Restricted Boltzmann machine: Recent
  advances and mean-field theory}},
  \href{https://doi.org/10.1088/1674-1056/abd160}{\emph{Chinese Physics B}
  {\bfseries 30} (2021) 040202}
  [\href{https://arxiv.org/abs/2011.11307}{{\ttfamily 2011.11307}}].

\bibitem{Aarts:2023uwt}
G.~Aarts, B.~Lucini and C.~Park, \emph{{Scalar field restricted Boltzmann
  machine as an ultraviolet regulator}},
  \href{https://doi.org/10.1103/PhysRevD.109.034521}{\emph{Phys. Rev. D}
  {\bfseries 109} (2024) 034521}
  [\href{https://arxiv.org/abs/2309.15002}{{\ttfamily 2309.15002}}].

\bibitem{Damgaard:1987rr}
P.H.~Damgaard and H.~H\"uffel, \emph{{Stochastic Quantization}},
  \href{https://doi.org/10.1016/0370-1573(87)90144-X}{\emph{Phys. Rept.}
  {\bfseries 152} (1987) 227}.

\bibitem{Verbaarschot:2000dy}
J.J.M.~Verbaarschot and T.~Wettig, \emph{{Random matrix theory and chiral
  symmetry in QCD}},
  \href{https://doi.org/10.1146/annurev.nucl.50.1.343}{\emph{Ann. Rev. Nucl.
  Part. Sci.} {\bfseries 50} (2000) 343}
  [\href{https://arxiv.org/abs/hep-ph/0003017}{{\ttfamily hep-ph/0003017}}].

\bibitem{AbulMagd:2014}
A.A.~Abul-Magd and A.Y.~Abul-Magd, \emph{Unfolding of the spectrum for chaotic
  and mixed systems},
  \href{https://doi.org/10.1016/j.physa.2013.11.012}{\emph{Physica A:
  Statistical Mechanics and its Applications} {\bfseries 396} (2014) 185–194}
  [\href{https://arxiv.org/abs/1311.2419}{{\ttfamily 1311.2419}}].

\bibitem{kingma:2017}
D.P.~Kingma and J.~Ba, \emph{{Adam: A Method for Stochastic Optimization}},
  \href{https://arxiv.org/abs/1412.6980}{{\ttfamily 1412.6980}}.

\bibitem{zenodo}
G.~Aarts, B.~Lucini, and C.~Park, chanjure/Stochastic\_weight\_matrix\_dynamics\_during\_learning \_and\_Dyson\_Brownian\_motion-data\_release: v1.0.1,
\href{https://doi.org/10.5281/zenodo.13294081}{{\ttfamily 10.5281/zenodo.13294081}}
 

\end{thebibliography}
\end{document}